\documentclass[12pt]{iopart}

\usepackage{graphicx}
\usepackage{verbatim}
\usepackage{lineno,hyperref}
\modulolinenumbers[5]
\usepackage{color}
\usepackage[export]{adjustbox}
\usepackage[]{subfigure}
\usepackage{placeins} 
\usepackage[english]{babel}  
\usepackage[utf8x]{inputenc}
\usepackage{booktabs}
\usepackage{tabularx}
\usepackage{multirow}
\usepackage{float}
\begin{document}
\title[Structural, Electronic and Mechanical properties of all-sp$^2$ graphene allotropes]{Structural, Electronic and Mechanical properties of all-sp$^2$ graphene allotropes: the specific strength of tilene parent is higher than that of graphene and flakene has the minimal density}

\author{Tommaso Morresi$^{1,2}$, Andrea Pedrielli$^{1,2}$, Silvio a Beccara $^1$, Ruggero Gabbrielli $^1$, Nicola M. Pugno$^{2,3,4}$,  and Simone Taioli$^{1,5}$}

\providecommand{\keywords}[1]{\textbf{Keywords:} #1}
\address{$^1$ European Centre for Theoretical Studies in Nuclear Physics and Related Areas (ECT*-FBK) and Trento Institute for Fundamental Physics and Applications (TIFPA-INFN), Trento, Italy}
\address{$^2$ Laboratory of Bio-Inspired and Graphene Nanomechanics - Department of Civil, Environmental and Mechanical Engineering, University of Trento, Italy}
\address{$^3$ School of Engineering and Materials Science, Materials Research Institute, Queen Mary University of London, UK}
\address{$^4$ Ket Lab, Edoardi Amaldi Foundation, Italian Space Agency, Italy}
\address{$^5$ Faculty of Mathematics and Physics, Charles University, Prague, Czech Republic}
\ead{taioli@ectstar.eu}
\ead{nicola.pugno@unitn.it}

\begin{abstract}
In this work a systematic approach to the search for all-$sp^2$ bonded carbon allotropes with low density is presented. In particular, we obtain a number of novel energetically stable crystal structures, whose arrangement is closely related to the topology of graphene, by modifying the packing of congruent discs under the condition of local stability. Our procedure starts from an initial parent topology and proceeds to generate daughter architectures derived by lowering the packing factors. Furthermore, we assess both the electronic properties, such as the band structure and the density of states, and the mechanical properties, such as the elastic constants and the stress--strain characteristics, of parent's and daughter's geometries from first-principle simulations. We find, using geometrical packing arguments, that some arrangements lead to a density as low as half that of graphene, obtaining some of the least dense structures of all-$sp^2$ bonded carbon allotropes that could ever be synthesized. Nevertheless, a threshold value of the density exists below which the mechanical rigidity of graphene is irreparably lost, while keeping other mechanical characteristics, such as the specific toughness and strength, almost unchanged with lower weight.
\end{abstract}

\vspace{2pc}
%\noindent{\it Keywords}: graphene allotropes with low density, augmentation, density functional theory, electronic and mechanical properties, stress-strain curves


\submitto{\TDM}
\maketitle
\section{Introduction}

Over the last decade graphene is arguably one of the most investigated materials among all the existing carbon allotropic forms \cite{doi:10.1002/anie.201600655,Geim2007}. Indeed, despite the difficulties in synthesising high-quality large-area graphene sheets \cite{taioli2014computational,tatti2016synthesis,taioli2016characterization}, its great promises and achievements in the fields of microelectronics, of materials science and chemistry motivate the great deal of scientific and technological efforts that scientists are pursuing.\\
\indent Further progress in the search of potential applications of graphene to various material/devices is mainly related to its unique electronic and mechanical properties \cite{doi:10.1063/1.4716178,taioli2009electronic,haberer2011direct,Allen2010,Randviir2014,Signetti2017,2053-1583-4-3-031013,PEDRIELLI2018766,AzzoliniJPCC,AzzoliniCarbon,PEDRIELLI2017796,haberer2010tunable}, even though the quest for practical applications have shifted the focus over the years to other layered materials, such as transition-metal dichalcogenides (TMDs) \cite{C4CS00182F}, silicene \cite{Vogt2012}, germanene, the monolayer form of black phosphorous \cite{Chen2016, Matthes2013}, and boron-nitride. Moreover, the power of modern supercomputer platforms, algorithms \cite{Marzari} and novel approaches e.g. based on artificial intelligence \cite{Nosengo} have paved the way also to the discovery of layered hybrid materials \cite{Geim2013} with the aim of combining together the most desirable characteristics of each layered structure \cite{Meyer2009,Signetti2017}, starting a new specific research field on 2D materials.\\
\indent Nevertheless, the possibility to introduce new interesting features also in bi-dimensional carbon-based materials without chemical functionalization, 
while keeping the desirable properties of graphene, such as its planar periodic structure and the $sp^2$ bonding network, might be very convenient to the existing technology. In this regard, one of the most striking properties of graphene is its Young's modulus to density ratio, probably the highest achieved so far. Unfortunately, investigations on this topic have been rarely pursued except for some notable exceptions \cite{Liu2012,Li-Chun2014,Wang2013,0953-8984-28-13-13LT01}. \\
\indent In this work we propose first a systematic approach for finding novel energetically stable structures characterized by $sp^2$-bonded carbon atoms of decreasing density using graphene as a frame of reference. In particular, we aim to find planar structures with density lower than graphene, possibly decreasing it up to the least dense form of carbon allotrope that could ever be synthesized, while displaying almost unchanged specific mechanical characteristics with respect to graphene. Indeed, one of the possible routes to increase the specific modulus with respect to graphene can be reducing the surface density. Increasing the specific modulus by decreasing the mass density is a typical request whereby the minimum structural weight can be achieved. This challenge has far-reaching consequences in a variety of applications, most notably in aerospace technologies where weight saving is a route to cost reduction. 
To characterize the response of these novel planar structures to external force and electromagnetic fields, we assess the stress--strain curves, their specific mechanical properties, and we calculate the electronic band structures of both the parent and derived daughter architectures from Density Functional Theory (DFT) simulations. Our analysis shows the existence of a threshold density below which the mechanical rigidity of graphene is very much depleted, while other specific mechanical characteristics, such as the strength and toughness, can be even bigger than graphene. \\
\indent Finally, we devise that the systematic approach presented in this work can be extended also to design novel lightweight strong three-dimensional carbon allotropes.

\section{Methods and computational details}\label{methods}

\subsection{Structure optimization.} The optimal structure search, electronic structure simulations and the assessment of the mechanical properties of the previously introduced architectures were performed within the DFT framework using the {\sc Quantum ESPRESSO} (QE) suite \cite{qe}. QE is a plane-wave code based on the pseudopotential approach to deal with the interaction between valence electrons and lattice ions. Optimization of the atomic configurations was carried out by using Broyden–Fletcher–Goldfarb–Shannon (BFGS) algorithm with the following DFT parameters. The simulation cells in the orthogonal direction to the plane of the structures is set to 20 \AA, in order to avoid spurious interactions among periodic images. The optimized configurations of all the structures investigated in this work can be found in the electronic supplementary material.
\subsection{Band structure simulations and DOS.} In our DFT simulations we used a norm conserving PBE pseudopotential and an energy cut-off for the wavefuntions equal to $100$ Ry. This large value of the plane-wave cut-off is due to obtaining converged values of the stress tensor, an observable notoriously more difficult to converge with respect to the total energy. The $k$-point grid used to calculate observables in the momentum space depends on the simulation cell size and was chosen so to achieve converged DFT values below chemical accuracy ($<$ 0.01 eV for the total energy and $< 10^{-3}$ Ry/\AA~ for the interatomic forces). Thus, depending on the simulation cell we performed calculations on $6\times 6 \times 1$ up to $16\times 16 \times 1$ $k$-point grids for structural minimization, while increasing the $k$-point mesh to $48 \times 48\times 1$ for the calculation of the Density Of States (DOS) and of the band structures. Convergence of the integrals over the Brillouin zone was improved by smearing the  occupancy with a $0.136$ eV width Gaussian function.\\
\subsection{Mechanical properties.} In order to properly sample the energy density function in linear regime for carrying out the calculation of the mechanical properties we used $0.001$-spaced points up to $0.01$ strain and further $0.005$-spaced points up to $0.05$ strain. To deal with the elastic deformations we used a supercell containing two unit cells for the materials having trigonal symmetry, i.e. the graphene and the flakene families. The $C_{11}$ coefficient in these cases is associated to the strain along the zig-zag direction. Upon deformation, the atomic positions within the supercell were relaxed until interatomic forces were smaller than $10^{-3}$ Ry/\AA. We further notice that in our ab-initio simulations we perform the calculations of the true stress. Thus, the simulation supercell is free to relax in the direction orthogonal to the loading below 0.5 kbar, equivalent to $0.5*20/3.35=3$ kbar. In fact, the calculation of the stress in the bi-dimensional material relies on the choice of a conventional thickness, which was set to 3.35 \AA~ for the graphene monolayer, while 20 \AA ~is the dimension of our simulation cell. The final pressure is calculated using the area resulting from relaxation, thus the plots of the stress-strain characteristics refer to the so-called ``true stress''.

\section{Results and discussion}

\subsection{Structure Search Method}

The structure of graphene-like materials is closely related to the packing of congruent discs touching each other exactly in three points. A two-dimensional packing can be achieved by a collection of congruent discs in the plane subject to the following constraints:

\begin{itemize}
    \item No two discs overlap;
    \item Each disc is in contact with at least another disc;
    \item For any choice of two discs in the packing there is always a path connecting them through mutual contacts.
\end{itemize}

Angle strain in $sp^2$ carbon allotropes increases noticeably when far from the equilibrium configuration equal to $2 \pi / 3$ rad. For this reason, we limit our study to structures containing only angles smaller than $\pi$ rad.
This angle choice corresponds to a specific condition for packing, called {\it local stability} or {\it locally jammed packing}. For locally stable disc packing, contacts between circles should lie not all on the same hemicircle \cite{Torquato2010}. With such a constraint in place, one might wonder whether packings of arbitrarily low density exist or, in case they do not, what the least dense arrangement of discs in the plane would be. It is an interesting question on its own, given that the question addressing the opposite problem $-$ that of finding the densest arrangement of discs in the plane $-$ received much attention for a long time \cite{Lagrange1773, Thue1910} and found a formal answer only in the last century \cite{Toth1943}. \\
In this regard, it is worth noticing that if the packing is allowed to be non-periodic, then discs can indeed be packed into locally stable configurations with arbitrarily low density \cite{Boroczky1964}.

\begin{figure*}
\centering
\includegraphics[width=1.0\textwidth]{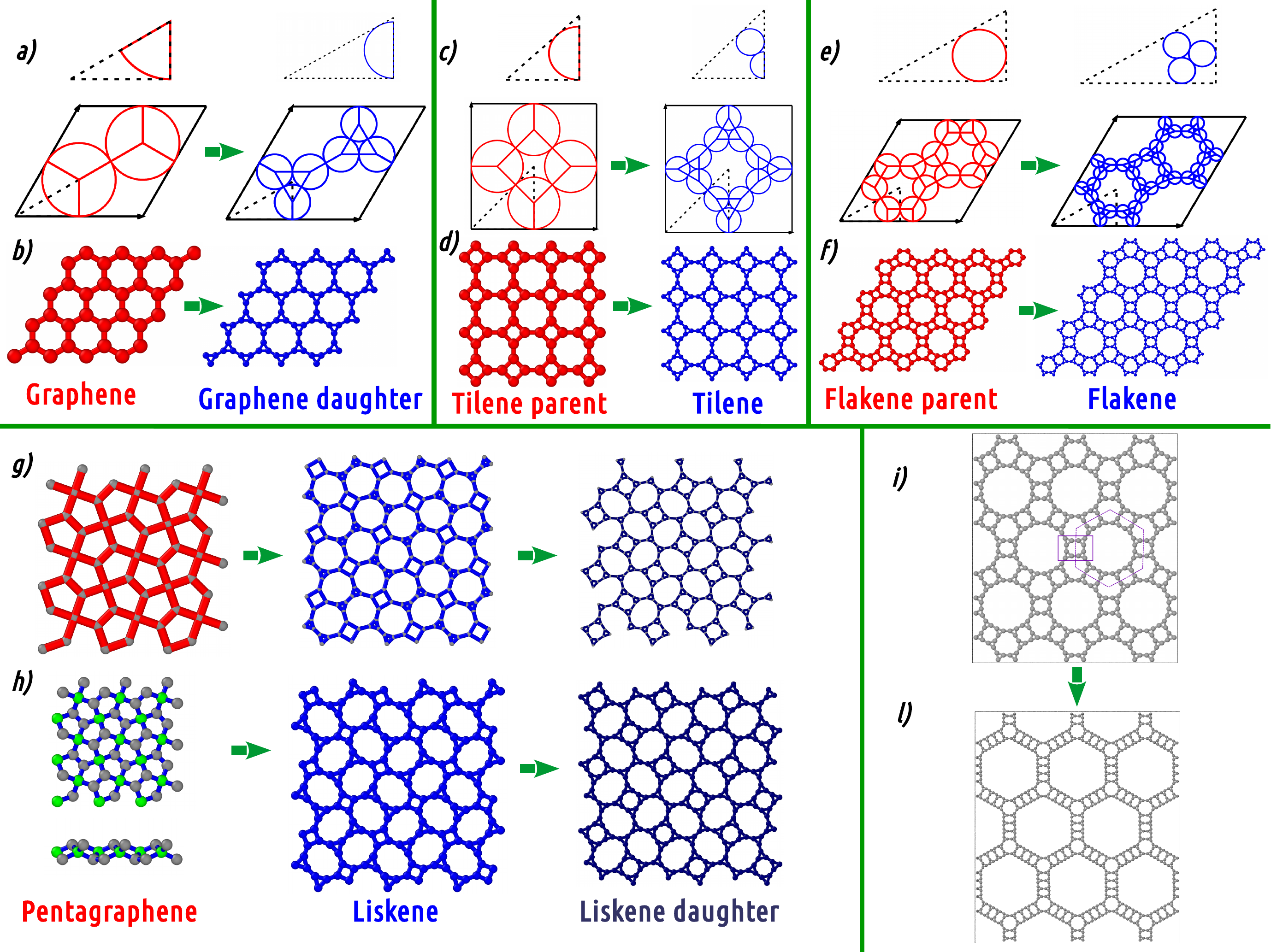}
\caption{First row: in the upper panel parent (left, red color) and daughter (right, blue color) disc packing in the unit cells of a) graphene, c) tilene, and e) flakene. The lines internal to the discs connect the nearest neighbor discs. Bottom panel: $4 \times 4$ supercells of parent (left) and daughter (right) structures of b) graphene, d) tilene, and f) flakene. Second row: 
g) from left to right: pentagraphene structure; augmentation of the Cairo pentagonal tiling: liskene; further augmentation of the liskene geometry: liskene daughter. h) from left to right: top and side view of a $3 \times 3$ pentagraphene super cell, where the $sp^3$-hybridized carbon atoms are reported in green color, while in grey scale we find the $sp^2$-hybridized carbon centers; $3 \times 3$ supercell of liskene and liskene daughter after performing DFT minimization. i) By relaxing the locally jammed packing constraint, the flakene structure can be made progressively less dense by elongating the hexagonal super-ring side highlighted in the picture. l) The low density structure obtained from flakene by doubling the hexagonal super-ring side length.}
\label{fig:structure}
\end{figure*}

\subsubsection{Graphene and graphene daughter.}

Consider the packing of discs associated with the structure of graphene. This arrangement has density of $\pi/(3 \sqrt{3}) \sim 0.6046$, as shown in the middle-left panel of figure \ref{fig:structure}(a), and defines the structural net of graphene reproduced in the left hand side of figure \ref{fig:structure}(b). Now replacing each disc with three smaller discs (reducing thus the disc radius by a factor $\frac{1}{1+2/ \sqrt{3}}$ with respect to the radius of graphene discs) - a process called augmentation when referred to nets \cite{Fischer2002} - leads to a less dense packing $\pi(7\sqrt{3})-12 \sim 0.390675$, as shown in the middle-right panel of figure \ref{fig:structure}(a). The resulting structure, which we name ``{\it graphene daughter}'' (gr11 in \cite{Sun2016}), is reproduced in the right hand side of figure \ref{fig:structure}(b). Unfortunately, the substitution cannot be pursued any further as contacts between circles on the half-hemicircle would occur and spoil the local stability condition. This applies to any packing, whose associated net (or tiling) contains triangles.

\subsubsection{Tilene parent and tilene.}

By considering tilings with polygons having a number of sides larger than three, in the middle-left panel of figure \ref{fig:structure}(c) we show what is probably the simplest structure after graphene, that is the packing associated with the well known square-octagon tiling. This defines a net of carbon atoms that we label ``{\it tilene parent}'' (octagraphene in \cite{Sun2016}), which is reported in the left hand side of figure \ref{fig:structure}(d). Its packing factor is $\pi (3-2 \sqrt{2}) \sim 0.539012$, which is lower than that of graphene. Its augmentation, carried out with the previously described principles, leads to an even rarer packing, as shown in the middle-right panel of figure \ref{fig:structure}(c), where the packing factor is $3 \pi / (2+\sqrt{2}+\sqrt{3})^2 \sim 0.355866$. The resulting structure, reported in the right hand side of figure \ref{fig:structure}(d), is called through this paper ``{\it tilene}''.

\subsubsection{Flakene parent and flakene.}

The tiling of the plane by regular polygons with the largest rings is the truncated trihexagonal tiling reported in the middle-left panel of figure \ref{fig:structure}(e) for which the packing factor before the geometry optimization is $\pi(2/\sqrt{3}-1) \sim 0.486006$. The resulting geometry is called ``{\it flakene parent}'' (C64 graphenylene in \cite{Sun2016}) and is reported in the left hand side of figure \ref{fig:structure}(f). Its augmentation, shown in the middle-right panel of figure \ref{fig:structure}(e), shows a tiling containing 24-sided polygons whose initial density is $3 \sqrt{3} \pi / (20+3\sqrt{3}+6\sqrt{7}+2\sqrt{21}) \sim 0.324951$. Although topologically equivalent, this packing differs from a previously reported example \cite{Fischer2002} as its density is appreciably lower. The resulting structure, obtained by placing carbon atoms at the disc centers, is reported in the right hand side of figure \ref{fig:structure}(f) and we call it ``{\it flakene}''. It can be shown that it is one of the $sp^2$ structures with lowest density ever studied which agree to the locally jammed packing conditions.

\subsubsection{Liskene.}

Other carbon structures can be designed by using a tiling conceptually different from what we have seen so far. In particular,
one may think to modify the carbon three-coordination, and to allow also four-coordinated vertices.
The Cairo pentagonal tiling is known to be the structure of pentagraphene \cite{Zhang2015} and it cannot be filled with congruent discs as the previously proposed structures. In this tiling indeed not all carbon atoms are three-coordinated and the resulting structure is not planar.
The unit cell of pentagraphene, whose 3$\times$3 periodic arrangement is reported in figure \ref{fig:structure}(g), is made by $4$ three-coordinated and $2$ four-coordinated vertices, characterized thus by $sp^2$-$sp^3$ and $sp^2$-$sp^2$ hybridization, respectively. This diversity of coordination leads the system to develop into the third dimension.
This reflects the fact that one cannot tile the plane by regular pentagons. Top and side views of the calculation supercell used in our simulation are reported in the left hand side of figure \ref{fig:structure}(h). To find the daughter structure, we apply our augmentation method also to the Cairo pentagonal tiling, characterizing the pentagraphene cell. In this way, we obtain a planar three-coordinated structure that we name ``{\it liskene}'', which is shown in the central panel of figure \ref{fig:structure}(g). While we notice that this is a different case study with respect to the other structures as the augmentation procedure takes place in a non-planar geometry, we find that the daughter architecture is still a three-coordinated system with a density lower than the parent. In the central panel of figure \ref{fig:structure}(h) we show the DFT optimized geometry of this tiling. Furthermore, by further augmenting liskene we obtain the daughter architecture (see the right panel of figure \ref{fig:structure}(g)), which represents the maximal limit of the planar packing of pentagraphene. Finally, in the right panel of figure \ref{fig:structure}(h) we report the DFT-optimized structure.

\subsubsection{Relaxing the locally jammed packing constraint.}

By relaxing the condition of having all the angles between two carbon bonds strictly less than $\pi$, the flakene architecture can be used as a basis for building up structures with arbitrary low density. This can be achieved by progressively elongating the sides of the hexagonal super-ring highlighted in figure \ref{fig:structure}(i), at fixed width. In particular, a sketch of the structure derived by doubling the hexagonal super-ring sides is reported in figure \ref{fig:structure}(l). Concerning stability, by increasing the length of the hexagonal sides to achieve an arbitrarily low density the energy-per-atom is $1.372$ eV/atom higher than graphene, which can be assumed to be the asymptotic value for area density going to zero.

\subsection{Structural optimization}

Prior carrying out the mechanical and electronic characterization of these novel carbon nets, we perform the structural optimization (details on the DFT parameters were given in section \ref{methods}).
In the second and third columns of Table \ref{tab2} we report the energy per atom and the cohesive energies obtained upon optimization of the atomic positions within the cell. \\
\indent The cohesive energy of graphene (7.74 eV) well agrees with the experimental value of 7.6 eV \cite{Dappe2006}, and with previous DFT simulations \cite{PASTI2018433} reporting a value of 7.828 eV. We notice that graphene is still the most energetically stable allotrope. In general, with the notable exception of pentagraphene, we observe that lowering the densities of the parent structures by using the previously introduced augmentation method results in daughter architectures characterized by lesser energetic stability and lower intra-molecular bond strengths. We rationalize the different finding in the case of pentagraphene, for which the cohesive energy increases from parent to daughter, by noticing that the augmentation starts from a non-planar $sp^2-sp^3$ net and ends up into a purely planar $sp^2$ net. This atomic arrangement represents thus a favourable solution from both the energetic and density points of view.\\
\indent For the flakene parent we find almost the same energy difference (0.6395 eV vs. 0.64 eV) with respect to graphene (see third column of table \ref{tab2}) as in \cite{Song2013}, where this structure is labelled ``graphenylene''. Also for the tilene parent we calculate an energy difference with respect to graphene equal to 0.5186 eV, which is very similar to the value of 0.53 eV reported in \cite{Liu2012}, where the structure was named T-graphene. Finally, in the case of pentagraphene we find an energy-per-atom difference of $0.904$ eV, which is very much comparable to the value of about $0.9$ eV reported in \cite{Zhang2015}.\\
\indent While we notice that the loss of stability is not significant, as the total energy difference per atom between the less stable material (flakene) and graphene is of the order of 1$\%$, the density is almost two times lower than that one of graphene (see the second column of table \ref{tab2}).
 
\begin{table*}
%\centering
\caption{\label{tab2}First column: structure type. Second column: surface density. Third and fourth columns report the total energy per atom with respect to graphene and the cohesive energy per atom obtained upon structural optimization, respectively. With the exception of pentagraphene, all structures are planar and each carbon atom is three-coordinated. In the table the following abbreviations were used: p.=parent, d.=daughter, dir.=direct gap, indir.=indirect gap.}
%\begin{indented}
\lineup
\begin{tabular*}{\textwidth}{@{}l*{15}{@{\extracolsep{0pt plus
12pt}}l}}
\br                          
%\begin{indented}
%\begin{tabular*}{|c|c|c|c|c|c|}
%\br
Structure & Density  & Energy  & Cohesive energy  &  Type & Bandgap\\
 &  (atoms/\AA$^2$) & ([eV]/atom) & ([eV]/atom) & & [eV]\\
\mr
\verb"Graphene"  & 0.379 &0 & 7.7404 & \verb"Semi-met." &  0 \verb"(dir.)"\\
\verb"Graphene d." & 0.256 & 0.9882 & 6.7523 & \verb"Metal" &  - \\
\verb"Tilene p." & 0.336 &  0.5186 & 7.2219 & \verb"Metal" &  -   \\
\verb"Tilene"  & 0.233 & 1.0765 & 6.6640 & \verb"Metal" & - \\
\verb"Flakene p." & 0.301 & 0.6395 & 7.1009 & \verb"Semi-met." &  0.043 \verb"(dir.)"\\
\verb"Flakene"  &  0.212 &1.1071 & 6.6334 & \verb"Metal" &  -\\
\verb"Pentagraphene" & 0.452 & 0.9044 & 6.8361 & \verb"Semicond" &  2.23 \verb"(ind.)"\\
\verb"Liskene"  & 0.297 & 0.7789 & 6.9615  & \verb"Semicond" &  0.36 \verb"(ind.)"\\
\verb"Liskene d." &  0.247 & 1.0506 & 6.6897 & \verb"Semicond" &  0.46 \verb"(ind.)" 
\\
\br
\end{tabular*}
%\end{indented}
\end{table*}

\begin{figure*}[!h]
\centering
\includegraphics[width=1.0\textwidth]{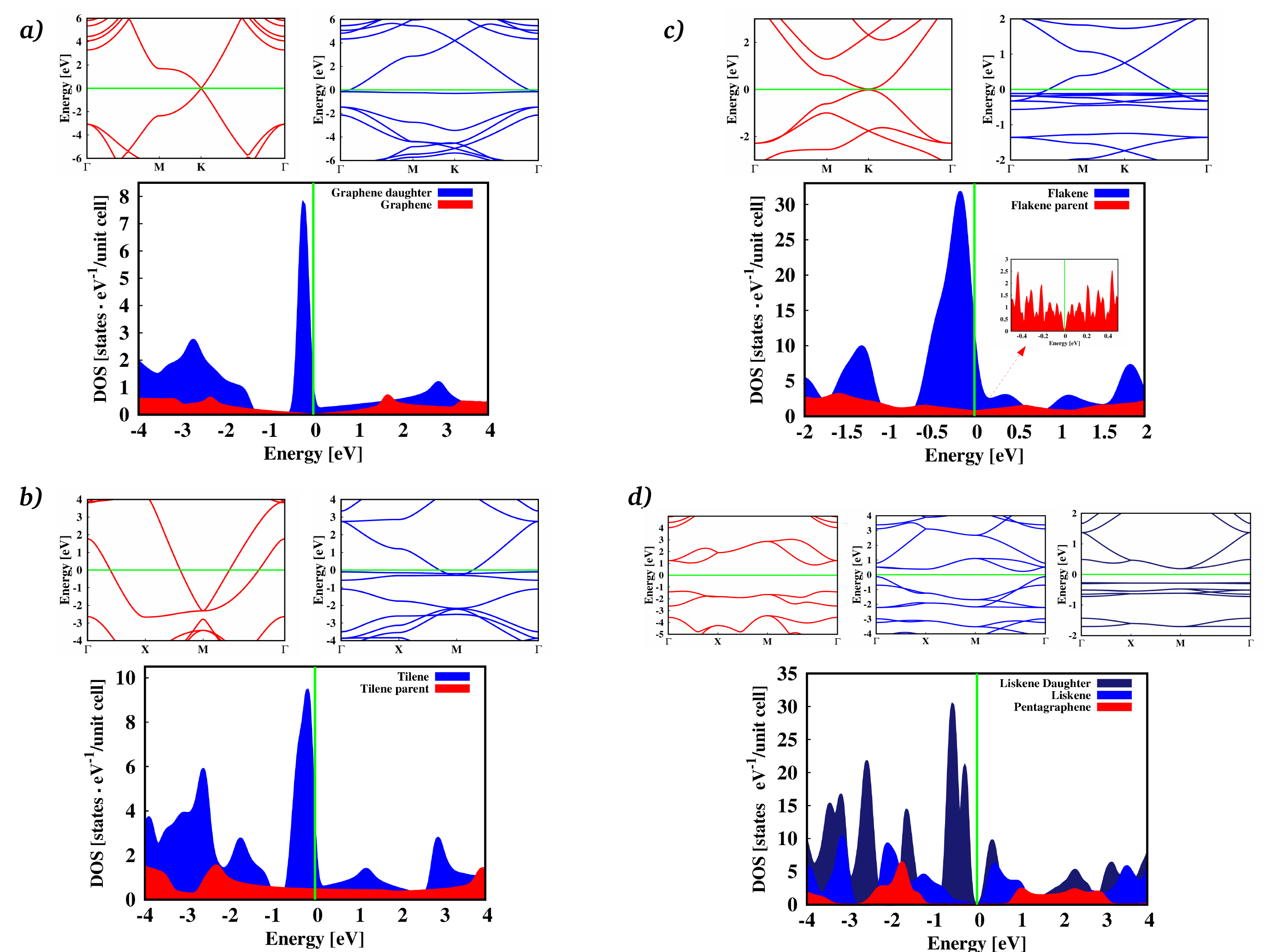}
\caption{Upper panels: band structure of the parent (left, red lines) and daughter (right, blue lines) architectures of a) graphene, b) tilene, c) flakene, and d) pentagraphene. Lower panels: relevant DOS of the parent (red filled area) and daughter (blue filled area) structures. Fermi level is shifted to zero and reported as an horizontal (vertical) green line in the band structure (DOS).}
\label{fig:dos}
\end{figure*}

\subsection{Electronic properties}

In this section we report the band structures and relevant DOSs for the eight structural arrangements proposed in this work. \\
\indent We begin with the well known electronic band structure of graphene reported in the upper left panel of figure \ref{fig:dos}(a) alongside the DOS (red filled curve in the lower panel of figure \ref{fig:dos}(a)), which we reproduce to test our choice of the DFT parameters. The agreement with previous simulations \cite{RevModPhys.81.109} is excellent so we can move to the assessment of the electronic structure of the other systems. \\
\indent In the right panel of figure \ref{fig:dos}(a) we report the bands of the graphene daughter. We observe that graphene loses its semi-metal characteristics and acquires a striking metallic behaviour with a loss of the typical graphene features near the Fermi energy (valence and conduction bands do not touch in one Dirac point as well as the dispersion around the $\Gamma$-point is not linear). This is due to the appearance of a narrow band close to the Fermi level (reported as an horizontal (vertical) green line in the band structure (DOS)), a feature that appears also in tilene and flakene, as can be seen in the upper right panels of figures \ref{fig:dos}b) and \ref{fig:dos}c). This is of course reflected into the DOS, characterized by a narrow peak close to the Fermi energy, as shown in the blue filled curves reported in the lower panels of figure \ref{fig:dos}(a), \ref{fig:dos}(b), and \ref{fig:dos}(c). \\
\indent Tilene parent and flakene parent electronic band structures (see upper left panels in figures \ref{fig:dos}(b) and \ref{fig:dos}(c), respectively) are not dramatically changed by augmentation, as the daughter structures stay metallic
(see lower panel in figure \ref{fig:dos}(b) for tilene daughter) or increase their metallic character (see lower panel of figure \ref{fig:dos}(c) for flakene daughter). \\
\indent At odds with the previous architectures, pentagraphene band structure (see upper left panel of figure \ref{fig:dos}(d)), which has the typical characteristics of a semiconductor (in agreement with previous DFT calculations \cite{Zhang2015}), is heavily affected by augmentation, as the daughter structure (see upper right panel of figure \ref{fig:dos}(d)) presents an almost semi-metallic behaviour characterized by a very narrow band gap. The relevant DOSs of pentagrafene (filled red curve of figure \ref{fig:dos}(d)) and liskene (filled blue curve of figure \ref{fig:dos}(d)) are typical of a semiconductor and semi-metal, respectively.

\subsection{Elastic properties}

\begin{table}
%\centering
\caption{\label{tab:Elastic} Elastic constants ($C_{11}$, $C_{12}$, $C_{44}$), area Young's modulus ($E_A$), Young's modulus ($E$), Poisson's ratio ($\nu$) and area specific Young's modulus ($E_A/\rho_A$) of the parent and daughter carbon structures. To evaluate the accuracy of our simulations, we report a comparison with data in the literature where available. In the table the following abbreviations were used: p.=parent, d.=daughter.}
\lineup
\begin{tabular*}{\textwidth}{@{}l*{15}{@{\extracolsep{0pt plus
12pt}}l}}
\br
          &   $C_{11}$  & $C_{12}$   & $C_{44}$   & $E_A$  & $E$ & $\nu$ & $E_A/\rho_{A}$ \\
          & (N/m) & (N/m)& (N/m)& (N/m)& (TPa) & & ($10^{-3}$ Nm~kg$^{-1}$) \\
\mr
\verb"Graphene"    & 348   & 53.8 & - &  340  & 1.14 &  0.154 & 1.79   \\ 
      \cite{Sun2016}  & 358  &  60 & - & 349 &   &   0.17&       \\
\mr
\verb"Graphene d."  & 149   &   94.0    & -     &  89.6 & 0.30 &  0.631  & 0.70 \\ 
     \cite{Sun2016}  & 152  &  98  & -  & 92.6  & & 0.64        &   \\
\mr
\verb"Tilene p." &  294  & 44.1 &  48.3  & 288 & 0.96 & 0.150 &   1.70   \\ 
         \cite{Lei2012}  &  296  &   46  & 49  & 306  & &  0.13     &     \\
\mr
\verb"Tilene"  & 124  & 75.5  &  11.3  & 78.6 & 0.26 & 0.607  & 0.67   \\
\mr
\verb"Flakene p."  & 220  & 57.7 &  -   & 205   & 0.69
 &  0.263   &  1.36 \\ 
\cite{Sun2016} & 227 &  61  & -  & 210  & & 0.27   &        \\
\mr
\verb"Flakene"  & 87.0   & 64.9  &  -  & 38.6  & 0.13 & 0.746  &   0.36 \\ 
\mr
\verb"Liskene"   & 187   & 94.8  &  52.0  & 138 & 0.46
 & 0.508  & 0.93 \\
 \mr
 \verb"Liskene d." & 127 &  65.6 &  19.4&  93.1
& 0.31 &  0.517
 &0.75\\
\br
\end{tabular*}
\end{table}

\begin{figure*}[!h]
\centering
\includegraphics[width=1.\textwidth,trim={0 0.2cm 0 0},clip=true]{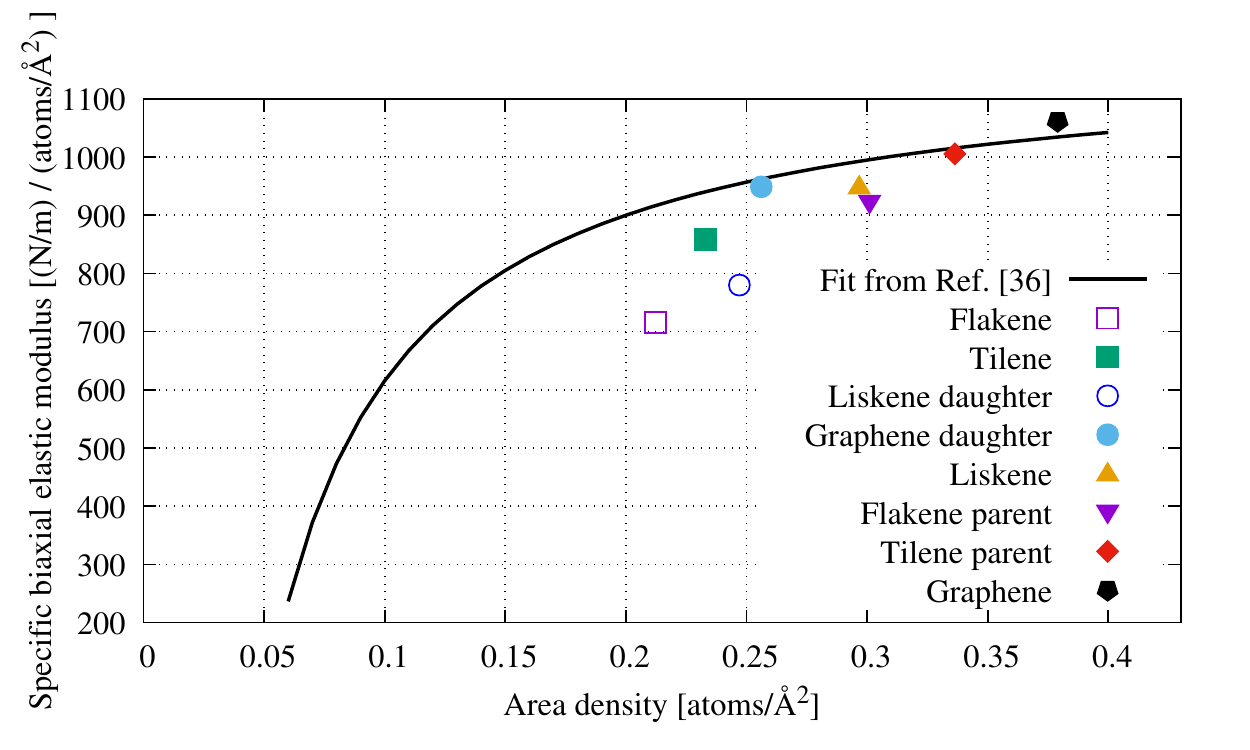}
\caption{Specific biaxial elastic modulus versus area density. The black line shows the trend reported in \cite{Sun2016} for carbon allotropes.}
\label{fig:Density}
\end{figure*}
To characterize the mechanical properties of the daughter architectures in comparison to the parent structures we carried out first the ab-initio simulations of the elastic stiffness tensor ${\bf C}$. The matrix ${\bf C}$ provides in linear approximation the proportionality or elastic constants relating the stress and the strain, $\bf{\sigma}=\bf{\varepsilon}{\bf C}$ where $\bf{\varepsilon}$ is the six-component strain vector, and $\bf{\sigma}$ is the stress tensor.
The stiffness tensor is in principle characterized by six independent terms in bi-dimensional materials, being $C_{ij} = C_{ji}$ for symmetry considerations. The elastic behavior of orthotropic 2D materials can be described thus by four elastic constants $C_{11}$, $C_{22}$, $C_{12}$ and $C_{44}$ \cite{Huntington1958}. For the square lattice structures, such as tilene parent, tilene and liskene, the symmetry constraint sets $C_{11}=C_{22}$, so that one has only three independent elastic constants. Graphene, graphene daughter, flakene parent and flakene at variance show isotropic hexagonal symmetry, reducing the independent elastic constants to only two according to the relations $C_{11}=C_{22}$ and $2C_{44}=C_{11}-C_{12}$.\\ 
\indent In harmonic approximation, the strain--energy density function $F$ at 0 K can be expressed as
\begin{equation}\label{enden}
    F=F_0+\frac{1}{2}F^{(2)}\varepsilon^2+o(\varepsilon^3)
\end{equation}
where $F_{0}$ and $1/2F^{(2)}\varepsilon^2$ are the static energy of the system and the lattice vibrational contribution, respectively. In our simulations we neglect the thermal electronic contribution, which is expected to be low.\\
\indent The elastic constants $C_{ij}$ can be then expressed as follows: 
\begin{equation}
   C_{ij}= \frac{\partial^2 F }{\partial \varepsilon_{i} \partial \varepsilon_{j}}
\end{equation}
The $C_{ij}$s can be derived by fitting the energy density of \ref{enden} with a second order polynomial in the imposed strain. In particular, on the one side for isotropic materials the fitting parameter $F^{(2)}$ can be identified with $C_{11}$ for uniaxial deformation and with $2(C_{11}+C_{12})$ under hydrostatic deformation, respectively. On the other side, in the case of orthotropic materials further calculations are needed in order to fully characterize the stiffness matrix. In this case, the elastic constants $C_{11}$, $C_{22}$ can be identified as the fitting parameters of the total energy under uniaxial strain, while $F^{(2)}$ corresponds to $C_{11}+C_{22}+2C_{12}$ or $4C_{44}$ in the case of hydrostatic deformation or shear deformation, respectively. 
From the knowledge of the elastic constants, the Young's modulus $E$, which measures material's stiffness, and the Poisson's ratio $\nu$, which measures the material's tendency to expand in directions perpendicular to the direction of compression, can be computed as $E=(C_{11}^2-C_{12}^2)/C_{11}$ and $\nu=C_{12}/C_{11}$, respectively.\\
\indent In table \ref{tab:Elastic} we report the elastic constants, Young's modulus and Poisson's ratio of all the 2D carbon allotropes studied in this work in comparison with the DFT values reported in the literature  \cite{Sun2016, Lei2012}, finding a remarkable agreement with previous calculations and experiments. We remind that, at variance with a stable, isotropic, linear elastic 3D material where the bounds on Poisson's ratio are $-1<\nu<1/2$, for 2D materials one has $-1<\nu<1$ \cite{Thorpe531}. Therefore, it is not surprising to obtain values of the Poisson's ratios higher than 1/2 for our 2D architectures. \\
\indent The Young's modulus of graphene obtained from our DFT simulations is in good agreement with the experimental value of 1$\pm$0.1 TPa (assuming a graphene thickness equal to 0.335 nm), obtained by nanoindentation measurements on single-layer graphene \cite{Lee385}. The analysis of the Poisson's ratio of tilene, flakene and liskene shows that these materials are almost incompressible. More precisely, the Poisson's ratio of tilene (as of graphene daughter, flakene, and liskene) is higher than the limit of isotropic incompressible 3D materials (which is $0.5$) while lower than the corresponding upper bound on Poisson's ratio for 2D materials (which is $1$ \cite{Thorpe531}): this material presents a hyper-restriction correspondent to a decrease of the area under tension. 
\begin{table}
\caption{\label{tab:stresstrain}Fracture strain (first column), strength (second column), strength$\times t$ (third column) and toughness$\times t$ (fourth column) of the parent and daughter planar structures alongside the specific strength and specific toughness (fifth and sixth columns). The conventional thickness of the graphenic materials is considered to be $t=3.35$~\AA. In the table the following abbreviations were used: p.=parent, d.=daughter.}
\lineup
\begin{tabular*}{\textwidth}{@{}l*{15}{@{\extracolsep{0pt plus
12pt}}l}}
\br
          & {\small Loading}  &  {\small Fracture}  & {\small Strength}   & {\small Strength } &  {\small Toughness}  & {\small Specific}   & {\small Specific}       \\
          & {\small direction} & {\small strain}  &    & {\small $ \times t$}  & {\small $ \times t$}  & {\small strength}  & {\small toughness}  \\
          & &  {\small (\%)} & {\small (GPa)}&  {\small (N/m)} & {\small (J m$^{-2}$)}& {\small (MNm kg$^{-1}$)}& {\small (MJ~kg$^{-1}$)}\\
\mr
\verb"Graphene"          & x &  $>$ 35 & 112 & 37.5 & $>$ 9.83 &  49.7 &  $>$ 13.0	 \\ 
                  & y & 26-28  & 102 & 34.2 & 6.51 &  45.2 &  8.61  \\ 
\mr
\verb"Graphene d." & x & 18-20  &  29.3   & 9.81 &  0.83    &  19.2 &  1.62  \\ 
                  & y &   $>$ 30 &   67.7  & 22.6 & $>$ 3.63   &  44.3 & $>$ 7.11   \\
\mr
\verb"Tilene p."     & x,y &  24-26  & 99.6  & 33.4 &  5.55  & 49.7 &  8.27 \\ 
       & $45 ^{\circ}$ & 32-34  &  79.1  & 26.5 &  5.68    &  39.5 &   8.47 \\
\mr
\verb"Tilene"            & x,y & 20-22  & 44.8 & 15.0 &   1.66 &  32.3 & 3.57  \\
      & $45 ^{\circ}$ & 18-20  &  30.3 & 10.2  & 0.92    &  21.9 &   1.97 \\
\mr
\verb"Flakene p."    & x &  22-24 & 66.7 & 22.3 &  3.37  & 37.2 &  5.61  \\
                  & y & 22-24   &  57.9 & 19.4  & 3.01    &  32.3  &   5.02 \\
\mr
\verb"Flakene"           & x & 12-14  & 23.6 & 7.92 &  0.49 & 18.7 & 1.16 \\ 
                  & y & 14-16   & 25.8  & 8.63 & 0.63   &  20.4 &  1.49 \\
\mr
\verb"Liskene"  & x,y & 18-20  & 63.2 & 21.2  & 2.19 & 35.8 & 3.70 \\ 
& $45 ^{\circ}$ & 14-16   & 43.8 &  14.7 & 1.27  &  24.7 &   2.14 \\
\mr
\verb"Liskene d." &  x,y &  12-14 &  27.8
& 9.32 & 0.59 & 18.5 & 1.20\\
 & $45 ^{\circ}$ & 14-16 & 28.0
&  9.37 & 0.68 & 19.0 & 1.37 \\
\br
\end{tabular*}
\end{table}

\begin{figure*}[!h]
\centering
\includegraphics[width=1.2\linewidth,trim={0 5cm 0 5cm},clip=true]{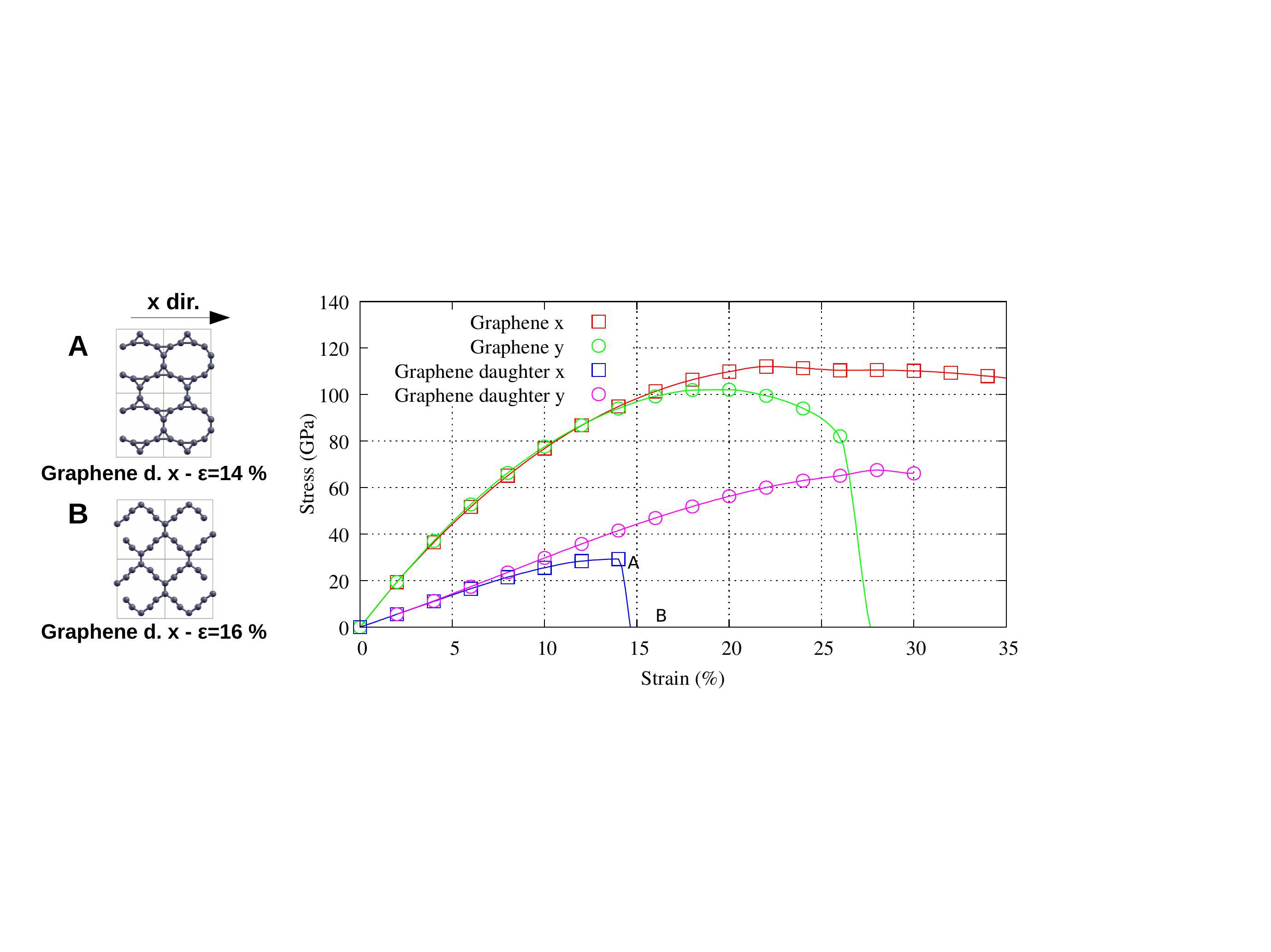}
\caption{Stress--strain curves of graphene and graphene daughter along the $x$-direction (or zig-zag, represented by red and blue empty squares for the two architectures, respectively) and the $y$-direction (or armchair, represented by green and violet empty circles for the two architectures, respectively). The differently colored lines represent the best fits to the ab-initio data. On the left and right sides of the image we report the simulation cells of graphene daughter for different strain values and directions.}
\label{fig:GrapheneFamily}
\end{figure*}

\begin{figure*}[!h]
\centering
\includegraphics[width=1.\linewidth,trim={0 5cm 0 5cm},clip=true]{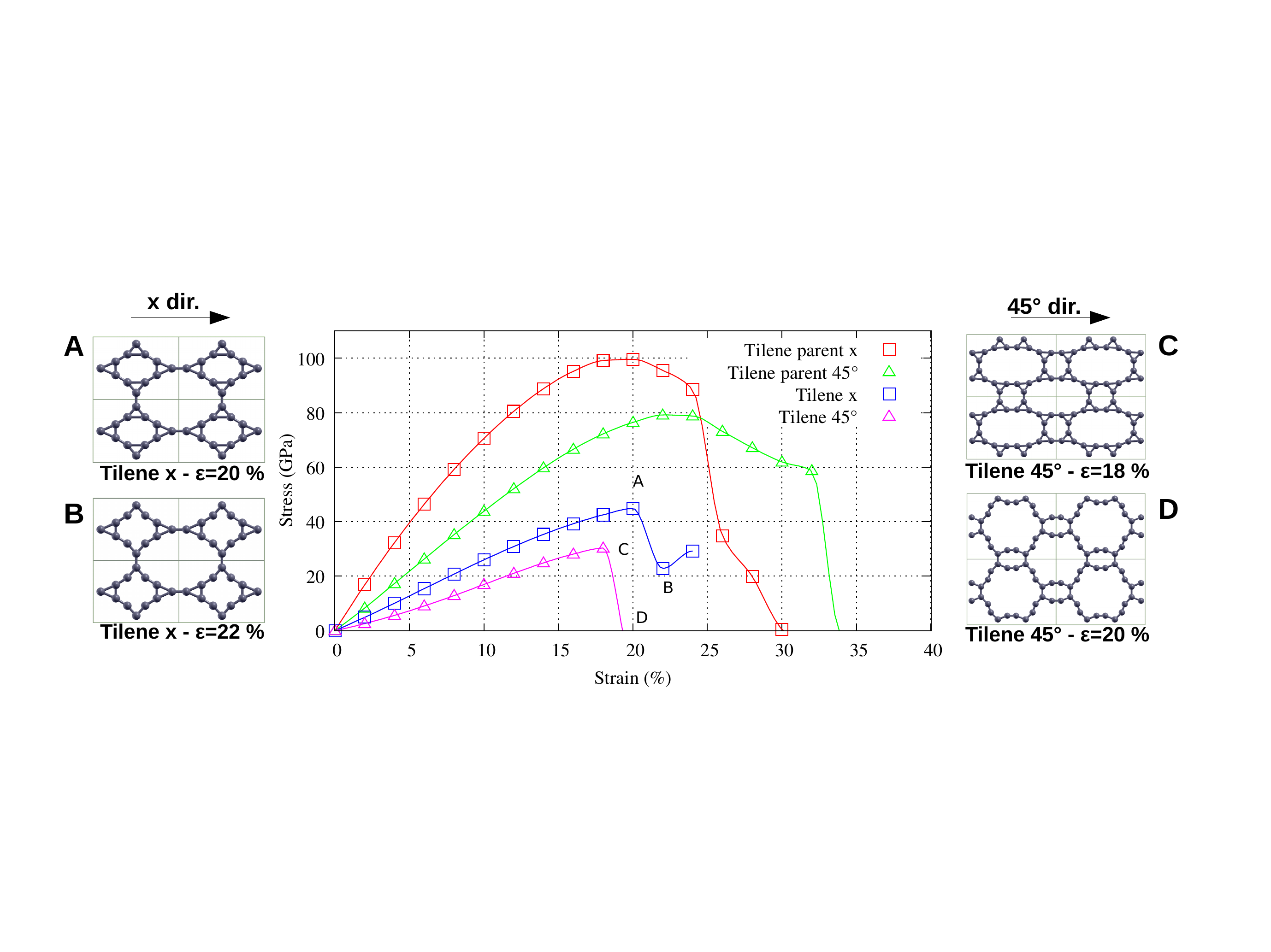}
\caption{Stress--strain curves of tilene parent and tilene along the $x$-direction (red and blue empty squares for the two architectures, respectively) and the $45 ^{\circ}$-direction (green and violet empty triangles for the two architectures, respectively). The differently colored lines represent the best fits to the ab-initio data. On the left and right sides of the image we report the simulation cells of tilene for different strain values and directions.}
\label{fig:TileneFamily}
\end{figure*}

\begin{figure*}[!h]
\centering
\includegraphics[width=1.\linewidth,trim={0 6cm 0 5cm},clip=true]{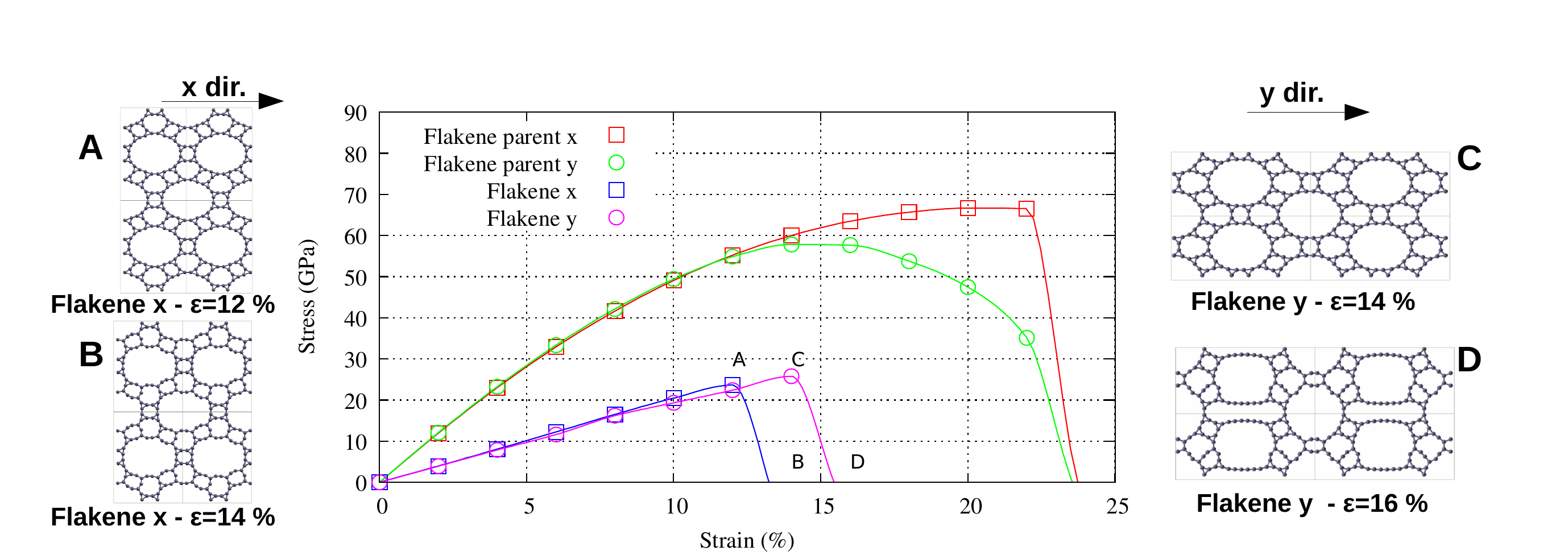}
\caption{Stress--strain curves of flakene parent and flakene along the $x$-direction (red and blue empty squares for the two architectures, respectively) and the $y$-direction (green and violet empty circles for the two architectures, respectively). The differently colored lines represent the best fits to the ab-initio data. On the left and right sides of the image we report the simulation cells of flakene for different strain values and directions.}
\label{fig:FlakeneFamily}
\end{figure*}

\begin{figure*}[!h]
\centering
\includegraphics[width=1.\linewidth,trim={0 5cm 0 5cm},clip=true]{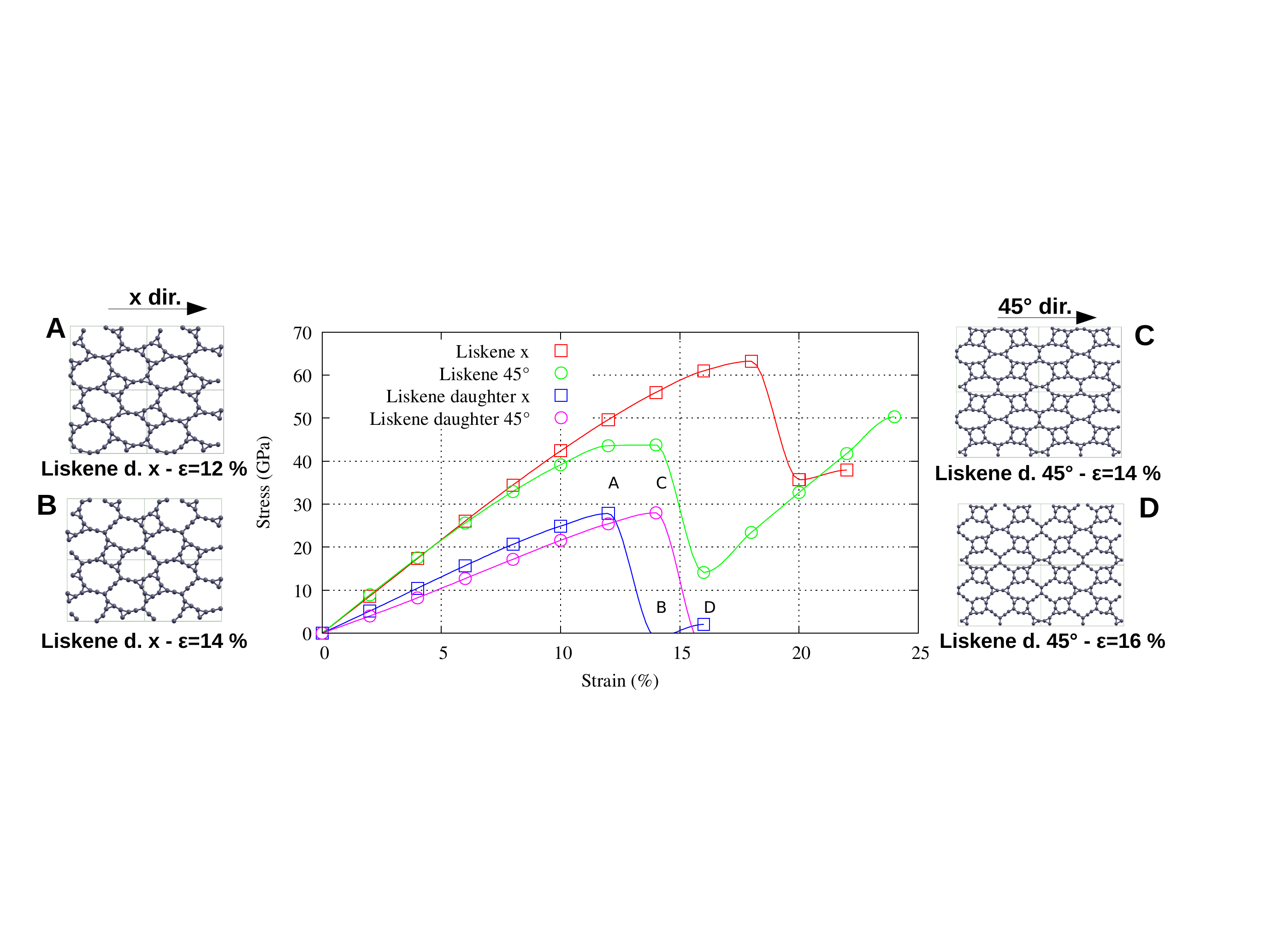}
\caption{Stress--strain curves of liskene and liskene daughter along the $x$-direction (red and blue empty squares for the two architectures, respectively) and the $45 ^{\circ}$-direction (green and violet empty circles for the two architectures, respectively). The differently colored lines represent the best fits to the ab-initio data.}
\label{fig:LiskeneFamily}
\end{figure*}

\begin{figure*}[!h]
\centering
 \includegraphics[width=1.\textwidth,trim={0 5cm 0 5cm},clip=true]{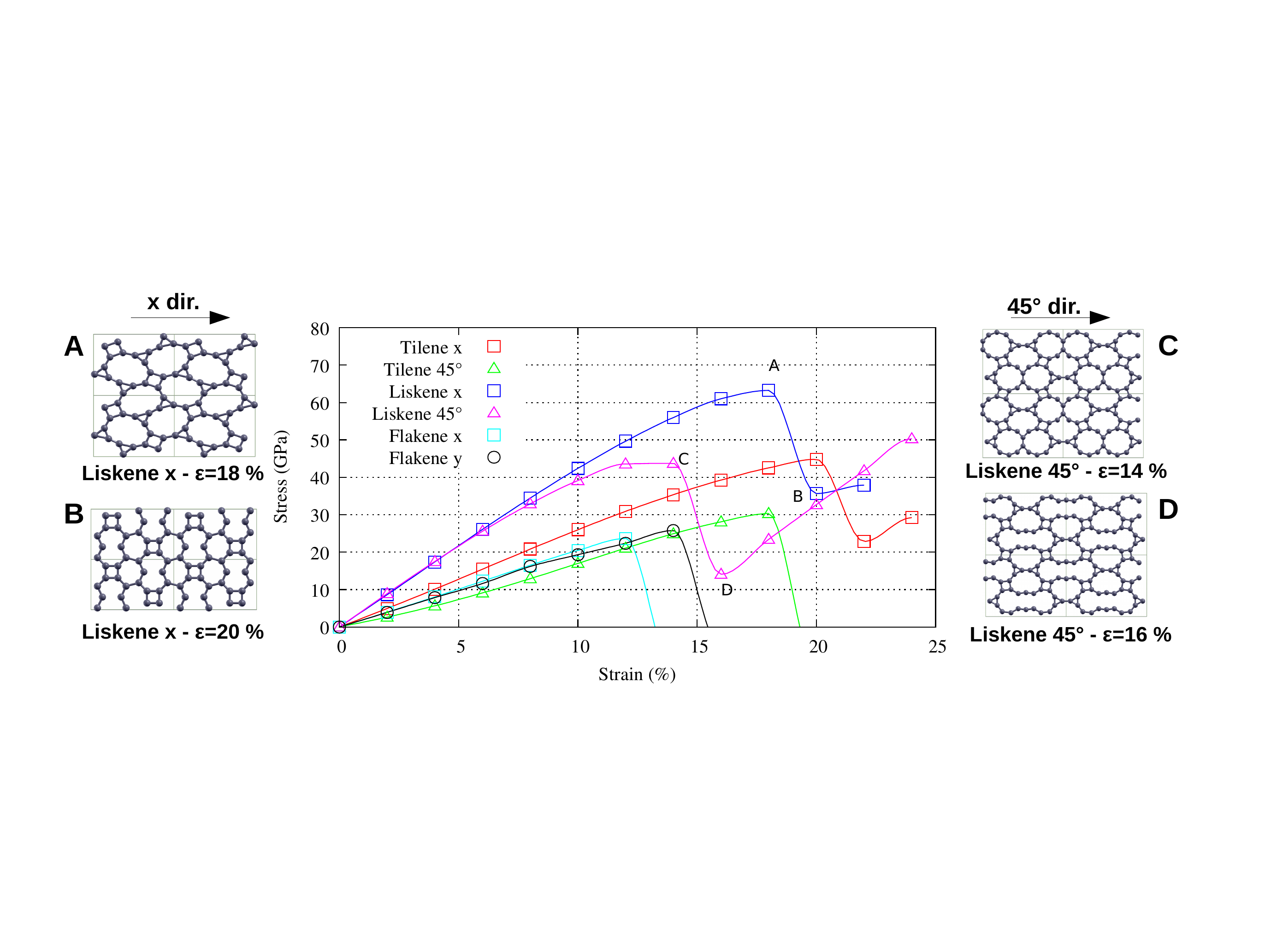}
\caption{Comparison between the stress--strain curves of liskene, tilene, and flakene along the $x$-direction (blue, red and cyan empty squares for the three architectures, respectively), along the $45 ^{\circ}$-direction (violet and green triangles for liskene and tilene, respectively) and along the $y$-direction (black empty circles for the flakene architecture). The differently colored lines represent the best fits to the ab-initio data. On the left and right sides of the image we report the simulation cells of liskene for different strain values and directions.}
\label{fig:NewStructures}
\end{figure*}

\begin{figure}[!h]
\centering
\includegraphics[width=1.\linewidth]{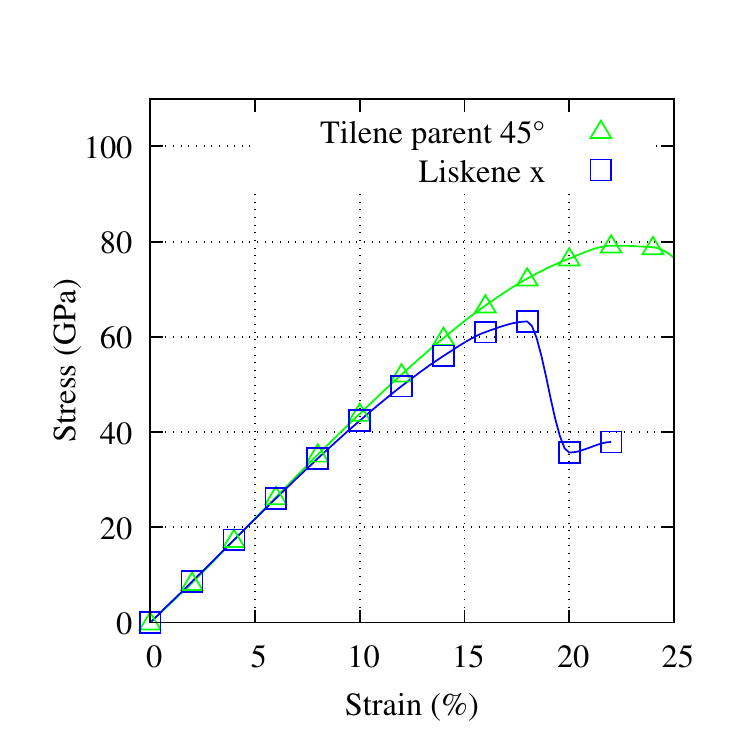}
\caption{Comparison between the stress--strain curves of liskene (blue empty squares) and tilene (green empty triangles) parent along the $x$ and $45 ^{\circ}$ directions, respectively. The differently colored lines represent the best fits to the ab-initio data.}
\label{fig:LiskenevsTilParent}
\end{figure}
Tilene presents an area Young's modulus $E_A=78.6$ N/m and a Poisson's ratio $\nu=0.607$, which are similar to those of the graphene daughter. Flakene has an area Young's modulus $E_A=38.6$ N/m and a Poisson's ratio $\nu=0.746$. Generally, we notice that graphene has the highest Young's modulus, and that moving from parent to daughter structures the Young's modulus decreases and the Poisson's ratio consequently increases. \\
\indent Moreover, one of the most significant observables to be computed for low density materials is of course the specific modulus, namely the Young's modulus divided by the mass density. Thus, we computed the Young's modulus per mass density $E_A/\rho_{A}=E/\rho$, where $\rho_{A}$ is the density in units of Kg/m$^2$ and $\rho$ the mass density in Kg/m$^3$. The outcome of our simulations concerning this quantity are reported in the last column of table \ref{tab:Elastic}.
We notice that graphene presents the biggest specific modulus among the materials studied here. At odds flakene, while displaying the lowest density among the investigated structures, shows a major drop in both the absolute and specific elastic moduli, which are from 8 to 5 times lower than graphene.
Nevertheless, while we do not find a material outperforming the specific properties of graphene in this respect and, thus, we do observe that the augmentation is only partially an advantageous route to follow in order to increase the specific modulus of graphene-like materials, the difference in the specific Young's modulus is less remarkable than for the absolute values, with the exception of flakene. \\
\indent The drop of flakene Young's modulus suggests that there is a threshold to the decrease of the density of these carbon-based planar materials, below which this mechanical characteristic is significantly depleted.
In order to get further insights on this issue, we report in figure \ref{fig:Density} the specific modulus of our carbon allotropes versus area density. In particular, we plot the specific biaxial modulus ($E_{bi}=C_{11}+C_{22}$) versus the area density, fitting the data reported in \cite{Sun2016} by the formula $E_{bi}=1184.3 \times \rho_A-56.88$ (N/m)/(atoms/\AA$^2$) (black curve in figure \ref{fig:Density}). We notice that the specific biaxial modulus of the structures studied in this work can be found in close proximity to the model fit.
These findings led us to the conclusion that the idea of decreasing the density, retaining the specific mechanical characteristics, can be pursued only to some extent at least as far as the Young's modulus is concerned. \\

\subsection{Stress--strain curves}

To gain further insight on the dependence of the mechanical properties of our structures on the density, we carried out the first-principles simulations of the stress--strain curves from which several observables can be obtained, such as the fracture strain, the tensile strength and the toughness. DFT calculations of the true stress tensor in response to strain, from which one can develop constitutive equations to fit the ab-initio data, were performed on the unit cell of the materials. We remind that all structures were relaxed below 3 kbar in the direction orthogonal to loading, and we plot the stress obtained by using the relaxed surface (true stress as opposed to engineering stress). \\
\indent In figure \ref{fig:GrapheneFamily} we start from analyzing the stress--strain curves of graphene and graphene daughter to benchmark our results against the extensive number of computational and theoretical studies carried out in this respect, along the Cartesian directions $x,y$, which represents the zig-zag and armchair directions of graphene (or, better to say, of the zig-zag and armchair ribbon that can be obtained by cleaving along the $x,y$ directions), respectively.
The stress--strain characteristics under uniaxial tensile loading along the zigzag ($x$, empty red squares) and armchair ($y$, empty green circles) directions, reported in figure \ref{fig:GrapheneFamily}, show the known anisotropic response of graphene that results in nonlinear constitutive equations \cite{XU20122582,doi:10.1021/jp307469u}.
The mechanical response of graphene to uniaxial tension is almost linear until about 10\% strain for both the armchair and zigzag directions, with the curve slope progressively decreasing with increasing strain. Beyond that value the stress--strain curves deviate significantly from linearity, keeping the isotropic behaviour up to $15$\% strain, where the mechanical characteristics along the two loading directions fork. The anisotropy develops at rather moderate strain with the zig-zag stiffness dramatically decreasing with respect to the armchair direction. In figure \ref{fig:GrapheneFamily} we sketch also the mechanical response to loading along the $x$ (empty blue squares) and $y$ (empty violet circles) directions of the graphene daughter. We notice that the absolute mechanical properties deplete significantly from graphene to its daughter in all respect, with a strong decrement in toughness and strengths (see also table \ref{tab:stresstrain}, where we report the absolute mechanical characteristics of these structures along with those of the other novel 2D architectures proposed in this work, that is tilene, flakene and liskene).
In this respect, we notice that the architecture of the graphene daughter is largely dominated by the presence of triangular shapes, at variance with graphene (see figure \ref{fig:structure}(b)). This feature is shared also by tilene (see figure \ref{fig:structure}(d)). This seems the major reason of the similar mechanical response to uniaxial strain along the $x$-direction between the graphene daughter (see violet empty circles in figure \ref{fig:GrapheneFamily}) and tilene (see blue empty squares in figure \ref{fig:TileneFamily} for a comparison). Along the $y$ direction the mechanical response of graphene daughter is similar to graphene (see violet empty circles in figure \ref{fig:GrapheneFamily}), showing a high fracture strain at lower stress than graphene.\\
\indent Tilene parent and tilene (see figure~\ref{fig:structure}(d)) belong to the dihedral group of symmetries (D4) and, thus, in figure \ref{fig:TileneFamily} we reproduced the stress--strain characteristics along the $x$ (empty squares, a strain along the $y$ direction would provide the same results) and the diagonal ($45^{\circ}$, empty triangles) directions.
Tilene parent displays a behavior under mechanical loading similar to graphene along the $x$-direction (empty red square in figure \ref{fig:TileneFamily}) with comparable strength and proportional limit stress (see table \ref{tab:stresstrain}). However, in the diagonal direction (see empty green triangles in figure \ref{fig:TileneFamily}) the presence of $sp^2$-carbon squares reduces the absolute mechanical performances of the tilene parent, but with a significantly higher fracture point (see table \ref{tab:stresstrain}). Nevertheless, the stress--strain characteristics do not overlap along the two different directions. Tilene shows a mechanical response to uniaxial strain comparable to graphene daughter in both the $x$ (blue empty squares of figure \ref{fig:TileneFamily}) and diagonal directions (violet empty triangles of figure \ref{fig:TileneFamily}), being its structure characterized by a similar occurrence of $sp^2$ triangles. \\
\indent Furthermore, in figure \ref{fig:FlakeneFamily} we report the stress--strain curves of flakene parent and daughter along the orthogonal directions $x$ (empty squares) and $y$ (empty circles) as their lattices display hexagonal symmetry. 
The mechanical characteristics of flakene parent are similar to those of graphene, showing a split between $x$ (empty red squares) and $y$ (empty green circles) curves at about $13$\% strain, and an almost linear regime up to $10$\% strain. 
However, the values of the strength are $60$\% lower than in the case of graphene (see table \ref{tab:stresstrain}). Flakene daughter shows a behaviour comparable to graphene daughter, being characterized by a similar large presence of $sp^2$-carbon triangular lattices, with lower absolute values of the strength (see table \ref{tab:stresstrain}).\\
\indent We notice that the augmentation procedure to obtain the liskene daughter architecture from liskene concerns only the carbon atoms that belong to the square shapes. In figure \ref{fig:LiskeneFamily} we report the stress--strain curves of liskene and liskene daughter along the orthogonal $x$-(empty squares) and $45^{\circ}$- directions (empty circles). Even in this case we find the general trend previously observed of a decrease of the fracture strain and tensile strength from the parent to the daughter structure.\\
\indent In figure \ref{fig:NewStructures} we present the stress--strain characteristics along the $x$ and $y$ directions of our novel planar architecture named liskene, compared to tilene and flakene.
As previously noticed, within the linear steep the uniaxial $x$ (blue empty squares) and $y$ (violet empty triangles) loading curves are overlapping, and at about 6\% strain they fork and deviate progressively from linearity up to the fracture strain at about 14\% and 18\% strain, respectively. The values of liskene mechanical characteristics are slightly higher than the other proposed architectures (see table \ref{tab:stresstrain}). We notice that after the disruption of the first set of bonds at 14\% and 17\% strain along the $x$ and $y$ directions respectively (see the structures reported in the left and right hand sides of \ref{fig:NewStructures}), liskene stress--strain curves in both directions bounce into a second linear regime with different slope. \\
\indent Finally, in figure \ref{fig:LiskenevsTilParent} we notice that tilene parent (green line) and liskene (blue line) basically present comparable stress--strain curves up to $10\%$ strain. We rationalize this by noting that for both structures the stress--strain characteristics are initially dominated by the deformation of $sp^2$-carbon atom arranged in square forms. However, as the strain increases, the liskene stress--strain curve departs from that one of tilene parent. This is due to the fact that the former architecture undergoes the fracture of the bonds within the squares, and the stronger bounds of carbon triangles come into play. \\
\indent As a final remark, we point out that the picture so far described concerning the absolute values slightly changes when we look at the specific properties reported in the last two columns of table \ref{tab:stresstrain}. Indeed, the strength of our novel 2D structures is comparable to graphene, or even higher than the latter in the case of the tilene parent. Nevertheless, the trend of the specific toughness, which measures the ability of a material to absorb energy before fracture, is generally favourable to graphene with respect to the other structures.

\section{Conclusion}

To conclude, in this work we present a systematic approach to the discovery of all-$sp^2$ carbon allotropes with the aim of decreasing the density of graphene without depleting its unique mechanical properties. This method proceeds by lowering the packing factors, which means augmenting the number of congruent discs under the constraint of local stability. \\
\indent While all the daughter structures that we generate display lower stability and smaller cohesive energy than graphene, their density is considerably lower than graphene up to $45\%$.
In particular, we argue that flakene represents the least dense possible structure among the families of all-$sp^2$ generated carbon allotropic forms starting from planar parent architectures under the local stability constraint. Nevertheless, we propose that novel geometries could be obtained by initiating the augmentation procedure from non-planar architectures, e.g. from pentagraphene. The relevant atomic arrangement derived from pentagraphene, which is named liskene, displays a high cohesive energy at a density lower than $22\%$  with respect to graphene. \\
\indent Nevertheless, by comparing the specific Young's modulus of these structures with graphene, we notice that there is a threshold below which is not possible to reduce further the density without a considerable depletion of this elastic property. In particular by lowering the density below that one of liskene results in a reduction of about $40\%$ of the specific Young's modulus. This can be clearly seen in the case of flakene, which displays the lowest density among the proposed planar structures as well as the smallest absolute and specific Young's modulus. Thus, graphene presents one of the highest specific modulus ever found and the quest for finding a better replacement in mechanical engineering applications is still open. Based on these findings, we argue that the focus on the search for materials with high specific Young's modulus should proceed among the high-density carbon allotropes. However, the area density of graphene is close to the limit of maximal planar packing. Furthermore, the specific Young's modulus has an asymptotic limit for high density packing. We further note that the structures with atomic density in the range of $0.25-0.3$ atoms/\AA$^2$ have performances similar to that of graphene and a research focused in this range of densities could be profitable in finding high specific modulus materials. Thus, a hypothetical improvement of this quantity could be devised only by changing the paradigm of interaction, for example by enhancing electrostatic and/or Van der Waals interactions. \\
\indent Our analysis of the mechanical properties concerned also the stress--strain curves of these low-density materials. We find that while the absolute values of the mechanical characteristics, such as fracture strain, strength, and toughness, are generally lower than those of graphene with the exception of the tilene parent architecture, nevertheless their specific counterparts can approach those of graphene and even surpass its specific strength for the case of tilene parent. In general, we notice that the mechanical properties deplete moving from parent to daughter architectures by lowering the packing factors. Thus, depending on the application, our structures could be used to replace graphene when weight decrease is an issue of paramount importance. \\
\indent Finally, we assessed also the electronic properties of the novel structures generated by our augmentation algorithm and compared them with their relevant parent networks. We find that a change in the packing factor results in the appearance of a narrow band close to the Fermi level, a common feature shared by all the parent-to-daughter architectures. This is particularly evident in the case of the liskene architecture, which is a semi-metal, despite the parent structure of pentagraphene is a semiconductor with a $2.3$~eV band gap according to our DFT simulations.\\
\indent We conclude by noticing that the systematic approach presented in this work could be extended also to design novel  lightweight strong three-dimensional carbon allotropes \cite{Qine1601536}.

\ack

N.P. is supported by the European Commission under the Graphene Flagship Core 2 grant No. 
785219 (WP14, "Composites") and FET Proactive ("Neurofibres") grant No. 732344 as well as by the Italian Ministry of Education, University and Research (MIUR) under the ``Departments of Excellence'' grant L.232/2016. The authors gratefully acknowledge the Gauss Centre for Supercomputing for funding this project by providing computing time on the GCS Supercomputer JUQUEEN at J{\"u}lich Supercomputing Centre (JSC) \cite{juqueen}. Furthermore, the authors acknowledge Bruno Kessler Foundation (FBK) for providing unlimited access to the KORE computing facility. 

%\FloatBarrier

\section*{References}
\bibliographystyle{unsrt.bst}

\end{document}